\documentclass{article}

\usepackage{arxiv}

\usepackage[utf8]{inputenc} 
\usepackage[T1]{fontenc}    
\usepackage{hyperref}       
\usepackage{url}            
\usepackage{booktabs}       
\usepackage{amsfonts}       
\usepackage{nicefrac}       
\usepackage{microtype}      
\usepackage{makecell}
\usepackage{siunitx}
\usepackage{caption}
\usepackage{subcaption}
\usepackage{multirow}
\usepackage{xr}
\usepackage{cleveref}
\usepackage{graphicx}

\pdfoutput=1

\title{A Multi-Platform Study of Crowd Signals Associated with Successful Online Fundraising}

\author{
  Henry K. Dambanemuya \\
  Northwestern University\\
  Evanston, IL 60208 \\
  \texttt{hdambane@u.northwestern.edu} \\
   \And
 Em\H oke-\'Agnes Horv\'at \\
  Northwestern University\\
  Evanston, IL 60208 \\
  \texttt{a-horvat@northwestern.edu } \\
}

\begin{document}
\maketitle

\begin{abstract}
The growing popularity of online fundraising (aka ``crowdfunding'') has attracted significant research on the subject. In contrast to previous studies that attempt to predict the success of crowdfunded projects based on specific characteristics of the projects and their creators, we present a more general approach that focuses on crowd dynamics and is robust to the particularities of different crowdfunding platforms. We rely on a multi-method analysis to investigate the \emph{correlates}, \emph{predictive importance}, and \emph{quasi-causal effects} of features that describe crowd dynamics in determining the success of crowdfunded projects. By applying a multi-method analysis to a study of fundraising in three different online markets, we uncover general crowd dynamics that ultimately decide which projects will succeed. In all analyses and across the three different platforms, we consistently find that funders' behavioural signals (1) are significantly correlated with fundraising success; (2) approximate fundraising outcomes better than the characteristics of projects and their creators such as credit grade, company valuation, and subject domain; and (3) have significant quasi-causal effects on fundraising outcomes while controlling for potentially confounding project variables. By showing that universal features deduced from crowd behaviour are predictive of fundraising success on different crowdfunding platforms, our work provides design-relevant insights about novel types of collective decision-making online. This research inspires thus potential ways to leverage cues from the crowd and catalyses research into crowd-aware system design.
\end{abstract}

\keywords{Fundraising, Crowdfunding, Collective Behaviour, Group Decision-Making}

\section{Introduction}
Increasingly, people recognise crowdfunding as an enabler of a variety of online fundraising activities that range from pro-social campaigns and supporting creative works to sizeable equity investments~\cite{zhang2012rational,agrawal2014some,belleflamme2014crowdfunding,althoff2014ask,mollick2014dynamics,gleasure2016emerging,vulkan2016equity}. This growing phenomenon is effective in reducing barriers in access to capital by eliminating the effects of geographic distance between project creators and funders~\cite{agrawal2011geography} and reducing the transaction costs of making such fundraising possible. Crowdfunding is also praised for promoting entrepreneurship by providing new opportunities to access funding~\cite{belleflamme2014crowdfunding,bruton2015new} and means to improve the livelihoods of people living in emerging economies~\cite{arvila2020enabling,bruton2015new}. In the wake of the recent novel Coronavirus pandemic, online fundraising has received heightened attention from many civic and international organisations that harnessed the power of crowdfunding to support their efforts due to a lack of traditional fundraising. For instance, the World Health Organisation (WHO) launched its first-ever crowdfunding campaign\footnote{COVID-19 Solidarity Response Fund for WHO:~\url{https://covid19responsefund.org/en/}} and several other eminent GoFundMe campaigns supported some of the most impacted countries, such as Italy\footnote{A record-breaking crowdfunding campaign is helping Italy fight Covid-19:~\url{https://qz.com/1836221/record-breaking-gofundme-campaigns-are-helping-italy-fight-covid-19/}}.

The growing popularity of online fundraising has attracted significant research on the subject. Most studies have tried to identify factors associated with successful fundraising, focusing on a \textit{single} platform (e.g.~\cite{agrawal2011geography,ceyhan2011dynamics,greenberg_learning_2014}), despite huge market variations both geographically and in the type of the fundraising effort. Existing research, therefore, often does not automatically generalise to other platforms and has resulted in conflicting findings concerning which project and creator determinants are associated with success. Furthermore, most prior studies have attempted to predict success based on various attributes of the projects~\cite{marom2013life,althoff2014ask,mitra2014language}, interactions with the crowd~\cite{xu2014show}, the creators~\cite{collier2010sending}, and their networks~\cite{freedman2008social,lu2014inferring,horvat2015network,ahlers2015signaling,vismara2016equity}. 

However, ad-hoc design and policy changes on crowdfunding platforms can confound all these factors~\cite{chakraborty2019impact}. Hence, the social computing community needs controlled approaches to systematically investigate the effects of project attributes and crowd behaviour on fundraising success. We thus present a general approach that is robust to the particularities of different crowdfunding platforms and markets and focuses on the crowd dynamics that contribute capital. This idea is backed up by evidence for the importance of successfully attracting funders early in the campaign~\cite{etter2013launch,colombo2015internal} and the role of subsequent herding in reaching the target amount~\cite{zhang2012rational,vismara2016information}. The broad spectrum of projects and creators, the quick pace of funding and untrained crowds using comparatively sparse data when selecting worthy projects are factors that substantially complicate decision-making in crowdfunding's low information and high-risk situations. In this context, most funders rely on collective cues when deciding to contribute to a project. Due to the significant signalling among crowd members, when and how much capital people provide becomes a crucial descriptor of the decision-making dynamics. Accordingly, previous work has found, on individual platforms, that simple features describing crowd dynamics can be significant markers of fundraising success~\cite{ceyhan2011dynamics,burtch2013empirical,agrawal2015crowdfunding}. We build on this observation by systematically investigating the dynamics of crowd behaviour across widely different crowdfunding platforms and markets through a multi-method analysis that relies on three different empirical methods to demonstrate the robustness of the crowd features. Our three main contributions are:

\begin{enumerate}
    \item We investigate similarities and differences between a charity platform that collects donations for public schools\footnote{\url{www.donorschoose.org}}, a dominant crowdfunding site that connects borrowers with lenders\footnote{\url{www.prosper.com}}, and a leading equity crowdfunding platform that offers investors the opportunity to buy shares in start-ups\footnote{We are unable to disclose the name of the platform due to our Non-Disclosure Agreement (NDA) with them.}. This is a unique multi-platform and cross-market study on crowdfunding success.
    \item We systematically test a set of intuitive and universal features that describe funder dynamics (\textit{crowd features}) and show their value in determining fundraising success within and across the studied platforms that span different markets, geographies, and fundraising efforts.
    \item To substantiate our analysis, we develop a framework that uses an innovative combination of methods for evaluating feature correlations and importance in a human-interpretable machine learning model as well as in matching samples along multiple dimensions to provide a causal understanding of the effect of crowd features.
\end{enumerate}

Our paper proceeds by first computing a set of crowd features that describe collective behaviour in a variety of settings that involve decision-making online. We first investigate correlation-based associations between individual crowd features and fundraising success. In combination with characteristics of projects that are visible to funders on each platform (\textit{project features}), we then perform supervised classification to predict fundraising outcomes and compare the predictive performance of crowd features to that of project features. Our results show that crowd features are significantly correlated with and better at approximating fundraising success across different online crowdfunding platforms than project features. However, since project features have been shown in prior research to determine fundraising success \cite{marom2013life,althoff2014ask,xu2014show,collier2010sending,emekter2015evaluating} and are observable to funders on the crowdfunding platforms, we rely on a quasi-experimental matching analysis to isolate and comparatively assess the effects of crowd features on fundraising success while controlling for the potential confounding influence of the observable project features. In particular, we use Coarsened Exact Matching (CEM)~\cite{iacus2012causal} to examine the causal effects of crowd features in relation to their specific crowdfunding platform settings and show that the crowd effects are robust to platform heterogeneity.

By demonstrating that universal features deduced from the behaviour of the contributing crowd are correlated with and predictive of fundraising success, even when controlling for project features observable by the crowd, our study provides empirical evidence of crowd dynamics features that are important in the funding success of projects across different platforms and robust to the particularities of the different online markets and platforms. Our work thus contributes not only to crowdfunding, crowdsourcing, and social computing literature but also to the growing body of knowledge on the science of success. We provide empirical insights on the emergence of crowd dynamics that eventually determine success in computer-supported cooperative work where collective cues underpin decision-making, thereby promoting research-based, crowd-aware platform design.

\section{Related Work: Dynamics of Crowdfunding}
Crowdfunding means raising money for a venture, cause, project, or organisation by drawing on relatively small contributions from a large group of individuals through a common online platform and without standard financial intermediaries~\cite{mollick2014dynamics}. Online crowdfunding emerged in the early 2000s through platforms such as DonorsChoose (2000), ArtistShare (2001), Prosper (2005), IndieGoGo (2007), and Kickstarter (2009). Since then, these platforms have attracted significant research attention in social computing and beyond (e.g.~\cite{azoulay2011incentives,gerber_crowdfunding_2013,althoff2014ask,mitra2014language,mollick2014dynamics,belleflamme2014crowdfunding,agrawal2014some,solomon2015don,dambanemuya2019harnessing}). Selecting from this vast literature, in this section, we discuss the current understanding on why project creators choose to crowdfund and what motivates diverse crowds to contribute towards crowdfunding projects. We further review the literature on known indicators of project success. We first focus on specific characteristics associated with successful fundraising and then detail findings that might generalise across different platforms.

For project creators, crowdfunding provides new opportunities to receive capital~\cite{gerber_crowdfunding_2013} especially for demographics with limited access to resources from traditional lending institutions~\cite{arvila2020enabling}. In the wake of the 2008 financial crisis, for example, crowdfunding became a viable solution for early-stage companies struggling to obtain funding through conventional financing~\cite{bruton2015new}. Project creators may also engage in crowdfunding for (1) establishing long-term interactions with funders that extend beyond the financial transaction and (2) receiving public validation for their projects and fundraising abilities~\cite{gerber_crowdfunding_2013}. Existing studies further show that crowdfunding platforms also range in terms of the motivations and goals of funders. For example, some funders are attracted to these platforms as a means of demonstrating their personal support to creators' projects~\cite{gerber_crowdfunding_2013}, in expectation of some kind of reward~\cite{gerber_crowdfunding_2013,belleflamme2014crowdfunding}, seeking to support an important cause with no expectations of reward~\cite{gerber_crowdfunding_2013}, or making a political statement\footnote{For instance via \url{www.crowdpac.com}}. Stark differences in motivations both for project creators and funders have given rise to various marketplaces and different crowdfunding models (e.g. lending, charity, equity, and reward-based crowdfunding). This heterogeneity in the nature of the fundraising effort raises the question: \textit{Which findings from individual platforms hold for crowdfunding in general?}

Despite the increasing public interest in crowdfunding, not all projects succeed. In fact, most projects fail to reach their fundraising goal by significant amounts and, typically, it is only by small margins that successful projects meet their goal~\cite{mollick2014dynamics,greenberg_learning_2014}. Identifying factors that lead to successful fundraising and predicting the probability of each project's success therefore remains one of the most important challenges in crowdfunding research. Several studies have linked fundraising success to the nature of the projects. For instance, across platforms like Kickstarter and Invesdor Oy (reward and equity platforms, respectively), the type of project matters because people tend to support efforts that reflect their cultural values or further causes they care deeply about~\cite{lukkarinen2016success,mollick2014dynamics}. As we would expect, the fundraising goal correlates with fundraising success as indicated by research on the reward-based platforms Kickstarter, Indiegogo, Ulule, Eppela, and Demohour. Specifically, projects that request large amounts of money are more likely to fail than modest requests~\cite{mollick2014dynamics,cumming2015crowdfunding,zheng2014role,cordova2015determinants}. Additionally, the framing of the request has also been linked to project success on the lending platform Prosper, on Kickstarter, and on the two charity platforms DonorsChoose and GoFundMe~\cite{larrimore2011peer,mitra2014language,rhue_emotional_2018}. Furthermore, according to research based on Kickstarter, Prosper, and AngelList (an equity platform) the visibility of the project helps with attracting funders. In particular, social media posts~\cite{etter2013launch,lu2014inferring,zhang_predicting_2017}, the size of the creators' social network~\cite{mollick2014dynamics,zheng2014role,greiner2009role,horvat2015network,hui2014understanding,chung2015long}, and their reputation~\cite{collier2010sending} increase chances of fundraising success. These studies indicate that various characteristics of projects, especially some that are specific to the platform, have an impact on potential funders' decision-making. 

There is a general consensus in crowdfunding literature that identifiable signals of quality play a key role in attracting contributions to projects. However, different platforms have different ways to signal project quality. For instance, project quality is often derived from descriptions that might include financial information, e.g. income statements may signal transparency, credibility, and feasibility~\cite{gafni2019life,lukkarinen2016success}. Additionally, media content on the fundraising page has also been linked with perceived project quality, mainly on Kickstarter. Particularly, a well-prepared concise video can quickly capture the attention of the audience~\cite{mitra2014language,mollick2014dynamics}, activity in terms of project updates might indicate productivity~\cite{xu2014show,lee_content-based_2018}, and funders' comments can suggest engagement and increase accountability among project creators~\cite{lee_content-based_2018}. Most importantly, research also supports that collective cues play a crucial role in funders' evaluation of individual projects. On the one hand, there is evidence for strong marketplace influences on funders' behaviour: other projects available on crowdfunding websites can draw money away~\cite{wash2014coordinating}, while the structure and design of the platform also affects crowd engagement~\cite{chakraborty2019impact}. On the other hand, in line with findings about the importance of information cascades and herding in successful fundraising~\cite{zhang2012rational,vismara2016information}, most funders interpret the amount~\cite{koning2013experimental} and arrival time~\cite{solomon2015don,colombo2015internal,agrawal2015crowdfunding,etter2013launch} of the first contributions as indicators of project quality. This crucial signalling among crowd members has triggered investigations into identifying descriptors of crowd dynamics that are associated with high-quality projects and successful crowdfunding~\cite{ceyhan2011dynamics,burtch2013empirical,agrawal2015crowdfunding,dambanemuya2019harnessing}. Yet, \textit{it remains unclear how important these crowd features are as determinants of success on different crowdfunding platforms after taking into account both general and platform-dependent project features.}

Existing research points to the need for a study that is based on multiple crowdfunding platforms as this might clarify contradictions in the literature about the importance of specific aspects either related to qualities of the project or the crowd dynamics among funders. For instance, a few studies have found a negative correlation between the duration of crowdfunding campaigns and their success~\cite{mollick2014dynamics,cumming2015crowdfunding,lukkarinen2016success}. While these studies suggest that longer fundraising campaigns may convey a message of indecisiveness and inability to deliver, Cordova et al.~\cite{cordova2015determinants} found that longer campaigns may also increase the likelihood of project success as the contributions will eventually add up to or even exceed the requested amount. Another example is the inconclusive finding about the role of activity on social media networks in fundraising success. Specifically, while some evidence suggests that project creators' social media posts are related to campaign success on Kickstarter~\cite{etter2013launch,lu2014inferring,zhang_predicting_2017}, research on Indiegogo, for instance, suggests otherwise~\cite{cumming2015crowdfunding}. Further work on Kickstarter observes that, although linked to the amount of early contributions, social media connections don't matter~\cite{colombo2015internal}. Possible explanations for the conflicting nature of evidence from these studies are that (1) they are based on different crowdfunding platforms and/or (2) different research methods were applied in each study. By conducting the same analysis on data from multiple crowdfunding platforms, we hope to resolve some of the contradictions in the literature and provide a robust assessment of the universality of crowd features.

\section{Data: Crowdfunding Platforms \& Markets}
We obtained data from three crowdfunding platforms that represent different markets both in terms of geography (US and UK) and the market model, i.e. lending, equity, and charity crowdfunding\footnote{Several studies have looked at reward-based crowdfunding, such as Kickstarter. Our analysis excludes the reward model due to the lack of fine-grained data about crowd dynamics on such platforms.}. These different platforms capture the heterogeneity in funders' motivations and goals which vary by the context and nature of the funding effort in each market model. For example, lenders and investors may be motivated by financial rewards~\cite{lu2012social,dezsHo2012lenders,ahlers2015signaling,cholakova2015does}, whereas donors on charity platforms may be motivated by reputation, self-image, or empathy-related rewards~\cite{choy2015affordances,gleasure2016does}. Additionally, the crowdfunding platforms differ in terms of their uses (e.g. paying for financial, entrepreneurial, or social ventures) and impacts (e.g. democratisation of financial services or greater availability of funding for pro-social projects)~\cite{gleasure2016emerging}. Across the different crowdfunding platforms, we further observe significant variation in the information that is visible to funders, for example, project details that inform potential contributors about the attributes of the project (e.g. auto loan, request for classroom book supplies, or business expense) as well as the characteristics of the project creators (e.g. their gender or income). Most notably, the data from the different platforms come from very different time periods (see Table~\ref{tab:summarystatistics}). The temporal component is further compounded by the fact that, at any considered time, different crowdfunding platforms and markets will be experiencing different levels of adoption and maturity. Considering the time differences across the platforms, a potential reliability of crowd dynamics features in consistently predicting project success would be unexpected and extremely interesting. Rather than provide a comparison between the different platforms, in this section, we introduce the three crowdfunding market models through representative platform data sets and describe important project variables that are available for prospective funders. In addition to identifying the project variables that are observable by funders on each platform, we further compute a set of variables deduced from the behaviour of the funding crowd and show in Section~\ref{sec:results} that features pertaining to crowd dynamics are significantly correlated with and predictive of fundraising success even after we control for the potential confounding influences of the observable project variables.

\paragraph{Lending Model} The Peer-to-Peer (P2P) lending model allows borrowers to receive varying amounts of commercial interest-based unsecured loans from crowd members~\cite{freedman2008social,greiner2009role,ceyhan2011dynamics,larrimore2011peer,horvat2015network}. The contributed funds are offered as a loan to be paid within a given time-frame and at a specified interest rate. We obtained crowd lending data from Prosper, the oldest P2P lending marketplace in the US. The lending data comprise 53,768 lenders who have collectively made 2,877,407 contributions towards 143,549 loans. The P2P platform attracts borrowers and lenders from all walks of life seeking non-collateral loans or small investments outside traditional financial institutions. For each project, the data describe characteristics of the loan, such as the requested \emph{amount}, \emph{interest rate} on loan, and \emph{monthly payment}. Included in the project information are attributes of the borrower, such as their \emph{Prosper score} i.e. a custom risk score built using historical Prosper.com data and allows the platform to maintain consistency when evaluating individual loan requests. There is also information about the \emph{credit grade} (i.e. the loan's estimated loss rate range), \emph{debt-to-income ratio}, and whether the borrower is a \emph{homeowner} or not. These project features are shown to lenders on the platform to signal each borrower's creditworthiness. Additionally, these features are commonly used by traditional financial institutions to make expert lending decisions based on borrowers' creditworthiness.

\paragraph{Equity Model} In equity crowdfunding, funders are investors entitled to shares of future profits in an entrepreneurial venture. Equity crowdfunding expanded rapidly after the 2008 financial crisis, but has grown slowly compared to peer-to-peer lending due to high levels of government regulation on securities as well as potential risks for fraud and the need for investor protection~\cite{bruton2015new,vulkan2016equity,lukkarinen2016success,vismara2016equity,vismara2016information}. We obtained equity crowdfunding data from one of the leading platforms in the UK and EU. The data comprise 21,907 investors who have collectively made 77,419 investments into 740 campaigns. On this platform, project creators include start-ups and early-stage companies seeking capital. Since projects are large capital campaigns, funders comprise both small and large institutional investors as well as wealthy individual investors. For each project, the data describe the requested \emph{amount}, \emph{equity percentage} offered in return of investment, and the company's \emph{valuation} prior to the investment. The data also describe the \emph{number of entrepreneurs}, whether the entrepreneurs have passed the finance \emph{quiz} to make sure that investors understand the risks of investing in startups and other growth-focused businesses, and whether the equity investment requires investor \emph{self–certification}, a process that requires investors to report their income and net worth as well as the amount of their other crowdfunding investments to reveal individual investor limits. Additionally, the project data describe whether the equity campaign is compliant with the UK's Enterprise Investment Scheme (\emph{EIS}) and Seed Enterprise Investment Scheme (\emph{SEIS}) which are tax incentive schemes for UK taxpayers who invest in qualifying early-stage businesses that are permanently established within the UK.

\paragraph{Charity Model} Some online fundraising efforts follow a charity model whereby funders serve as philanthropists who expect no material or financial return for their donations~\cite{koning2013experimental,althoff2014ask,wash2014coordinating,solomon2015don}. We obtained charity crowdfunding data from DonorsChoose, one of the earliest crowdfunding platforms that allows individuals to make donations towards public school classroom projects. The charity data comprise 850,498 donors who have made 1,004,658 donations to 215,825 public school projects from pre-K to grade 12. The projects are posted by teachers from different parts of the US and from communities in rural, urban, and suburban areas. They span several subject areas from math and science to literacy and language. For each project, the data describe the requested \emph{amount}, teacher's \emph{gender}, students' \emph{grade level}, \emph{community type}, \emph{subject area}, and the \emph{type of resource} that the donations are intended for (e.g. books, technology equipment, art supplies, or school trips). Similar to the other platforms we study, these project details are visible to donors (i.e. funders) on the site.

The notable differences between these crowdfunding markets and platforms are reflected in the different project features listed above. From the different project features observable on each platform, funders then decide what projects to support based on their expectations of each project's success deduced from the project variables that they believe to be associated with success. However, these project variables do not capture the role of funders' contribution patterns towards project success~\cite{zhang2012rational,vismara2016information}. In the next section, we therefore describe the crowd features that characterise funders' behaviour across all of these platforms and provide details about our methods for (1) investigating the relationship between the crowd features and fundraising success, (2) predicting fundraising success and comparing the relative importance of project and crowd features in the predictive task, and (3) estimating the quasi-causal effects of the crowd features on fundraising success.

\section{Predicting Successful Fundraising}
On all three platforms, we only considered projects that were either fully funded, or failed to meet their funding goal. We excluded active projects, DonorsChoose projects that received funds re-allocated from failed projects as these projects did not reflect true funder activity, as well as Prosper projects that had no credit information\footnote{The platform stopped showing borrowers' credit grade to funders in 2009 and hence we focus on projects posted before that time. Throughout our analyses, credit grade is an important variable of creditworthiness because this is the most common indicator of financial health used by lenders in traditional financial settings.}. Table~\ref{tab:summarystatistics} provides a high level summary of the data. 

\begin{table*}[ht]
\centering
\caption{Summary of our data collected from three different crowdfunding platforms. As shown, data were collected across multiple years, but at different times for the three platforms. The crowdfunding platforms also differ in terms of the number of projects, contributors, and contributions (i.e. loans, investments, and donations made to various projects). Bottom row summarises computed crowd features (mean, std) for each platform.}
\label{tab:summarystatistics}
\begin{tabular}{|l|l|l|l|}
\hline
 Variable & Lending & Equity & Charity \\ \hline
 Period & 2005 - 2008 & 2013 - 2015 & 2002 - 2016 \\
 Projects & 143,549 & 740 & 215,825 \\
 Contributors & 53,768 & 21,907 & 850,498 \\
 Contributions & 2,877,407 & 77,419 & 1,004,658 \\ \hline
 Appeal & 19.041 (40.318) & 104.620 (175.694) & 4.655 (4.906) \\
 Momentum & 1.100 (0.876) & 1.080 (0.505) & 1.023 (0.595) \\
 Variation & 0.384 (0.513) & 2.416 (1.854) & 0.516 (0.495) \\
 Latency & 0.458 (0.419) & 0.289 (0.324) & 0.616 (0.236) \\
 Engagement & 7.029 (2.221) & 52.449 (38.681) & 33.557 (44.384) \\ \hline
\end{tabular}
\vspace{-.3cm}
\end{table*}

\subsection{Crowd Determinants of Fundraising Success}
In addition to the project features identified above, we computed general crowd features that characterise the collective dynamics of fundraising that ultimately decide what is worthy of success. In contrast with old theories claiming that genius and personal performance are behind outstanding achievements in science, technology, business, and the arts~\cite{simonton1999origins, merton1968social, bowler2010making, sternberg1988nature,james2009economy}, there is increasing evidence for the collective nature of success~\cite{barabasi2005network, guimera2005team}. Within this new line of research, there is indication that the crowd-based valuation process is to a great extent random~\cite{salganik2006experimental} and that arbitrary initial advantages are inflated by positive feedback. We believe this collective aspect can help us navigate the increasing number and diversity of indicators conceivable and available via Web-based platforms to approximate fundraising success via the broad appeal, crowd engagement, as well as the variation and temporal patterns in fundraising activity. We therefore compute the following five crowd features based on arguments from prior literature:
\begin{itemize}
    \item Intuitively, the more funders a project attracts, the more likely that it will meet its funding goal. Hence, we count the number of unique funders of each project and consider that to be the project's \emph{appeal}. We expect appeal to correlate with success as it has been shown to in previous studies~\cite{belleflamme2014crowdfunding,gerber_crowdfunding_2013}.
    \item Temporal aspects of funders' activity, such as the arrival times of individual contributions might also signal confidence in the project's merit~\cite{ceyhan2011dynamics}. Accordingly, our next feature focuses on the speed at which funds are accumulating, as a reflection of how fast funders make their determination to contribute. We measure the \emph{momentum} of contributions through the coefficient of variation for the times between consecutive contributions i.e. the ratio between the mean and standard deviation of these time intervals.
    \item Along a similar argument, we also measure the \emph{variation} in contribution amounts using a coefficient of variation. The main idea here is that the amount of others' contributions visible to funders can influence also the behaviour of the crowd~\cite{burtch2013empirical}. This feature signals potential herding mechanisms that have been found to influence contribution dynamics on lending platforms~\cite{ceyhan2011dynamics}.
    \item Further, prior work has also found that early contributions to crowdfunding may signal the crowd’s interest in a project thereby attracting other funders to contribute as well~\cite{solomon2015don}. To measure this temporal aspect, we compute each project's \emph{latency} as the difference between the time of the first contribution and the time that the project was posted.
    \item Finally, for each project, we compute a crowd \emph{engagement} feature as the time between the first and last contribution when the project reached either its fundraising deadline or goal. While in some cases this measure may correlate with project duration, it captures only the time frame in which funders were actively contributing to a given project.
\end{itemize}

For all projects on the three crowdfunding platforms, we computed these five features. Summary statistics per platform are shown in Table~\ref{tab:summarystatistics}.

\subsection{Methods}
\label{sec:methods}
In this section, we introduce the methods that make up our multi-method analysis. We used Pearson's correlation to investigate the relationship between crowd features and crowdfunding success. We then combined crowd features with project features provided by each platform to train and evaluate the performance of Random Forest classifiers in predicting fundraising success~\cite{breiman2001random}. Essentially, these were binary classifications aiming to differentiate between funded and failed projects based on available features. Since the range of values for each feature vary wildly, we use min-max normalisation to scale the features to a fixed range from 0 to 1. The result of this pre-processing technique is that each feature contributes approximately equally to the learning process and hence the model's sensitivity decreases due to the relative scales of features. We also tried other classification methods such as Logistic Regression, Naive Bayes, and Adaptive Boosting. The results with these alternative methods were qualitatively indistinguishable from the ones obtained with Random Forest, which are also interpretable and allow for a better understanding of feature importance, which becomes crucial when comparing the relative importance of crowd features to that of project features. Since crowd lending and charity platforms have a large class imbalance (20.2\% and 99.4\% funded projects, respectively), we under-sampled the majority class and performed the classification task on balanced data on all platforms. In all experimental setups, we perform $k$-fold cross validation with hold-out samples. Specifically, for each platform, we randomly divide the data into $k=5$ subsets. Each time, one of the $k$ subsets is used as the test set (hold-out samples) and the other $k-1$ subsets are combined to form a training set. Then we compute the average accuracy, precision, recall, F1-Score, and area under the receiver operating characteristic curve (AUC) across all $k$ trials. We further evaluated the importance of individual and grouped (project vs crowd) features in predicting fundraising success using the Random Forest permutation importance (piRF) score which is measured as the relative increase in the model's prediction error after permuting the individual or grouped features' value. We rely on Scikit-Learn's Python API for the Random Forest implementation~\cite{scikit-learn}.

While Random Forest permutation importance scores provide a systematic ranking of crowd and project features based on how predictive they are of fundraising success, they cannot help understand \emph{why crowdfunded projects with similar covariates sometimes end up with dissimilar outcomes} or \emph{identify differences in crowd behaviour that may explain such seemingly arbitrary outcomes}. To investigate this question, we rely on Coarsened Exact Matching (CEM) which is a widely-used method for deriving causal inferences from observational data where the treatment variable is not randomly assigned~\cite{iacus2012causal}. Specifically, CEM provides a quasi-experimental approach for assessing the effects of crowd dynamics features on fundraising success while controlling for the confounding influence of project features that are associated with funding success. Common in the social sciences, this method has been used effectively to investigate the effect of race in online dating~\cite{lewis2013limits}, the impact of temperature and precipitation variability on the risk of violence in sub-Saharan Africa~\cite{o2014effects}, and the influence of women's inner social circles on their leadership success~\cite{yang2019network}. 

The CEM approach begins by identifying and grouping projects with similar platform-specific features observable by funders, but with varying crowd features. Lending crowdfunding projects were matched based on the requested amount, monthly loan payment, interest rate, Prosper score, credit grade, debt-to-income ratio, and homeownership. Equity projects were matched according to the requested amount, equity percentage offered, pre-money valuation, number of entrepreneurs, investor self-certification and quiz status, and EIS and SEIS compliance. Charity projects were matched based on the requested amount, resource type, teacher's gender, students' grade level, subject area, and community type. We then rely on CEM's automated algorithm for ``coarsening'' these project features to discrete values or  ``bins'' and matching projects with exact  ``bin signatures'' thereby generating groups of similar projects.

We categorised projects into treatment and control groups based on whether they were successfully funded or not, then estimated the effect of each crowd feature on fundraising outcome (i.e. fully funded or not), while controlling for project features. We do so using the traditional CEM measure of Sample Average Treatment Effect on the Treated (SATT) measure: $SATT = \frac{1}{n(T)} \sum\limits_{i \in T} \{(Y_{i}|T_{i}=1) - (Y_{i}|T_{i}=0)\}$ where $Y_i$ is the outcome variable (funded ($Y_i = 1$) or not ($Y_i=0$)), $T$ is the set of crowd treatments ($T_1$=Appeal, $T_2$=Momentum, $T_3$=Variation, $T_4$=latency, $T_5$=Engagement), and $n(T)$ is the number of crowd treatment effects, i.e, five. We thus compute the sample average treatment effect of each crowd feature on fundraising success as the difference between two possible outcomes. For each project, the \emph{fundraising outcome under crowd treatment condition} $(Y_{i}|T_{i}=1)$ is always observed. However, the counterfactual condition $(Y_{i}|T_{i}=0)$, i.e. the \emph{fundraising outcome if no treatment condition}, e.g. if no crowd appeal, momentum, variation etc., is always unobserved and imputed via simulation using a logit model. Once the unobserved outcomes are imputed, the estimate of each crowd feature's sample average treatment effect is measured by simply averaging the differences over all observations and imputed outcomes for the counterfactuals $(Y_{i}|T_{i}=1) - (Y_{i}|T_{i}=0)$. The SATT therefore follows the Rubin causal model (RCM), an approach to the statistical analysis of cause and effect based on the framework of potential outcomes~\cite{holland1986statistics}. Based on the RCM, the causal effect of each crowd feature is therefore the difference in fundraising outcome between the observed and counterfactual condition.

To allow for comparisons with other matching methods that retain all treated projects and select an equal number of control projects to include in the matched data set based on a distance or similarity measure, e.g. nearest neighbour matching (NNM), we further pruned the CEM solution using the Euclidean distance within each matched sample to achieve similar one-to-one matching solutions with CEM as one would obtain with NNM. In this case, the advantage of CEM over other matching methods is that for each project in the treatment group (funded = 1) we have exactly one ``twin-project'' in the control group (funded = 0) that has the exact same coarsened project features as the project in the treatment condition. Any projects in the treatment group that have no ``twin-project'' are thus discarded. This additional filtering procedure ensures that we are making counterfactual inferences only from valid points of comparison~\cite{king2011comparative,king2017balance}. To assess the goodness of the matching solutions, we used the $L1$ statistic ($1$: perfect imbalance, $0$: perfect balance) which is a measure of global imbalance with respect to the joint distribution of the project covariates. The $L1$ statistic is not valuable on its own, but serves rather as a point of comparison between matching solutions, thus $L1$ works for imbalance as $R^2$ works for model fit: the absolute values mean less than comparisons between matching solutions~\cite{blackwell2009cem}. In comparison to nearest neighbour matching, CEM produced better matching solutions and hence provides a more reliable approach for deriving causal inferences from the observational data used in this study.

\section{Results}
\label{sec:results}
\begin{figure}[!hbp]
    \centering
    \includegraphics[scale=.3]{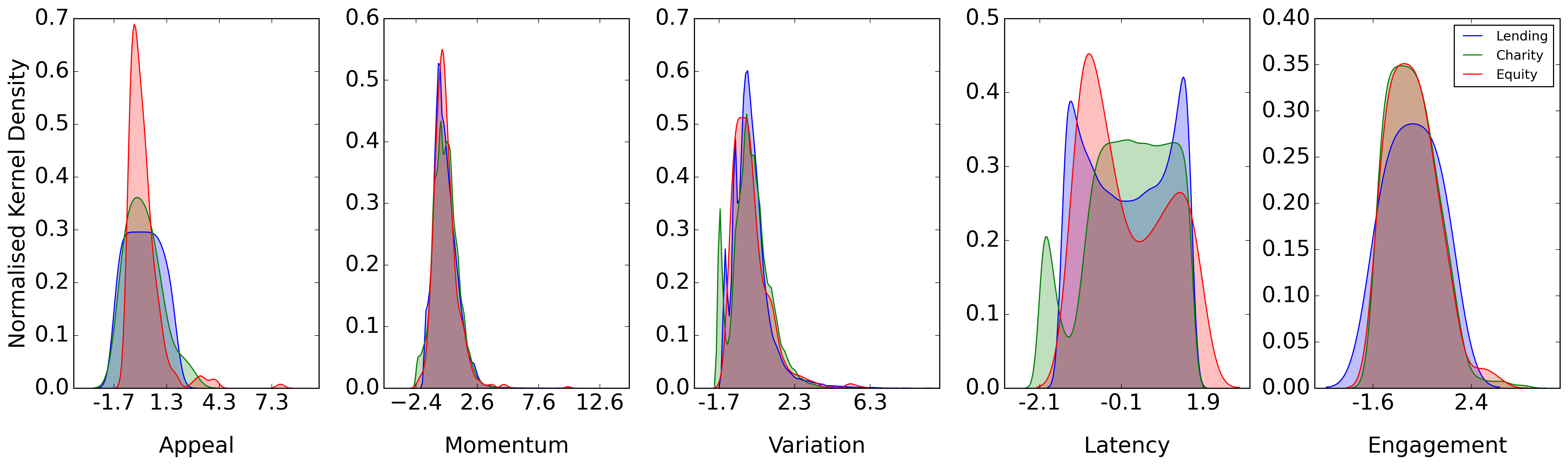}
    \caption{A comparison of kernel density estimates of crowd features on different crowdfunding platforms shows similar distributions that describe the underlying behaviour of funders on each platform.}
    \label{fig:kde}
\end{figure}

We observe similar crowd behaviour across the different crowdfunding platforms, despite differences in the number of projects posted per unit time, individual contribution amounts towards each project, and project funding success rate on each platform. The kernel density estimates of the crowd features on all three platforms share similar distribution properties indicating similarities in crowd activity in terms of individuals' underlying decisions about whether or not to fund a project, how quickly the crowd decides to fund a project, how quickly funds are accumulating, variation in contribution amounts, and how long funders remain engaged in fundraising (Figure~\ref{fig:kde}).

We further empirically test the degree of multimodality of the crowd feature distributions using Hartigan's Dip Statistic (HDS)~\cite{hartigan1985dip} and observe that crowd appeal, momentum, and engagement follow uni-modal distributions (Dip test: $p=1.0$). The crowd's latency follows a bi-modal distribution (Dip test: $p<0.05$) whereby some projects receive a substantial number of contributions early, while other projects take much longer to secure those initial contributions. The shapes of the bi-modal distributions also resemble the ``bathtub'' effect (named after its shape), which is most notable on the lending platform. This effect has been observed in simulation studies of funders' donations over time on the Donors Choose platform~\cite{solomon2015don}. The ``bathtub'' effect in crowd latency arises when projects either quickly receive funds immediately after being posted or go through an initial period of few to no contributions due to lack of crowd appeal or funders choosing to observe other people's contributions before making their own. 

\subsection{Crowd Features are Correlated with Fundraising Success}

On all platforms, statistical comparisons between the mean values of crowd features for funded and failed projects show that successful projects have greater appeal, higher momentum of contribution activity, and greater variation in contribution amounts compared to failed projects (Table~\ref{tab:crowdstatisticsbyoutcome}). Thus the crowd's appeal, momentum, and variation in contribution amounts are significantly positively correlated with fundraising success on all crowdfunding platforms. These findings support previous qualitative and quantitative findings that demonstrated the role of the number of contributors and frequency in contributions on fundraising success~\cite{belleflamme2014crowdfunding,gerber_crowdfunding_2013,ceyhan2011dynamics,colombo2015internal}. Our results also lend empirical evidence to qualitative studies as they show that the higher the variation in contribution amounts, hence less herding in funders' contributions, the more likely a project is to reach its fundraising goal~\cite{bikhchandani1998learning,lu2012social}. Based on these findings, we therefore anticipate that for crowdfunded projects to be successful, they need to appeal to all sorts of funders, big and small, whose contributions complement each other to meet the fundraising goal.

\begin{table*}[!h]
\centering
\caption{Mean (std) values of crowd features by project category and funding outcome. Pearson correlation between crowd features and fundraising success. Accordingly, crowd feature values are statistically significantly different for funded and failed projects. The only exception is latency on the charity platform. Notation: * significant at $p < 0.05$; ** significant at $p < 0.01$; *** significant at $p < 0.001$.}
\setlength{\tabcolsep}{1pt}
\label{tab:crowdstatisticsbyoutcome}
\resizebox{\columnwidth}{!}{%
\begin{tabular}{l|l|l|l|l|l|l|l|l|l|}
\cline{2-10}
 & \multicolumn{3}{c|}{Lending} & \multicolumn{3}{c|}{Equity} & \multicolumn{3}{c|}{Charity} \\ \cline{2-10} 
 & \multicolumn{1}{c|}{\makecell{Funded \\ 20.2\%}} & \multicolumn{1}{c|}{\makecell{Failed \\ 79.8\%}} & \multicolumn{1}{c|}{$r$} & \multicolumn{1}{c|}{\makecell{Funded \\ 35.3\%}} & \multicolumn{1}{c|}{\makecell{Failed \\ {64.7\%}}} & \multicolumn{1}{c|}{$r$} & \multicolumn{1}{c|}{\makecell{Funded \\ 99.4\%}} & \multicolumn{1}{c|}{\makecell{Failed \\ 0.6\%}} & \multicolumn{1}{c|}{$r$} \\ \hline
\multicolumn{1}{|l|}{Appeal} & 67.544 (62.957) & 6.754 (16.924) & 0.605*** & 175.789 (174.561) & 31.399 (43.619) & 0.534*** & 3.951 (3.953) & 2.204 (1.976) & 0.038*** \\
\multicolumn{1}{|l|}{Momentum} & 1.906 (0.784) & 0.759 (0.664) & 0.599*** & 1.422 (0.534) & 0.881 (0.360) & 0.518*** & 1.025 (0.595) & 0.636 (0.544) & 0.040*** \\
\multicolumn{1}{|l|}{Variation} & 0.946 (.0588) & 0.242 (0.377) & 0.551*** & 3.511 (2.042) & 1.819 (1.427) & 0.436*** & 0.517 (0.495) & 0.303 (0.425) & 0.033*** \\
\multicolumn{1}{|l|}{Latency} & 0.135 (0.256) & 0.539 (0.413) & -0.388*** & 0.208 (0.310) & 0.332 (0.323) & -0.184*** & 0.616 (0.236) & 0.605 (0.233) & 0.004 \\
\multicolumn{1}{|l|}{Engagement} & 5.762 (3.020) & 7.350 (1.833) & -0.287*** & 57.180 (37.646) & 49.871 (31.170) & 0.104** & 33.352 (44.093) & 83.833 (60.817) & -0.088*** \\ \hline
\end{tabular}
}
\vspace{-.1cm}
\end{table*}

We further anticipate that funders are more likely to contribute to projects with notable initial contributions compared to projects with little to no initial contributions. This hypothesis is based on previous research that shows that while projects with a moderate-sized initial contribution slightly outperform projects with no contribution, small initial contributions significantly decrease the chances of success for a project~\cite{koning2013experimental}. On the lending and equity platforms, we observe that the shorter the crowd latency (i.e. first funders respond quickly to a posted project) the more likely a project will reach its fundraising goal, hence significant negative correlations. This finding supports previous qualitative studies that highlight the importance of early donations in making the fundraising goal easier to achieve by reducing the remaining funds needed, while at the same time signalling project quality and funders' buy-in and decisiveness on a project's merits~\cite{solomon2015don,mollick2014dynamics,cumming2015crowdfunding,lukkarinen2016success}. We observe no significant correlation between crowd latency and fundraising success on the equity platform. Finally, we observe that crowd engagement is significantly negatively correlated with fundraising success in the lending and charity platforms meaning that successful campaigns typically take less time to be fully funded compared to those that are unlikely to succeed. In contrast to relatively small contributions on lending and charity platforms, we anticipate that equity campaigns targeting large contributions require significantly more fundraising time and effort to reach full funding. Our expectations are confirmed and projects do need more engagement to reach the investment goal on the equity platform.

\subsection{Crowd Features Predict Fundraising Success Better than Project Features}

We further combined the crowd features with project features provided by each platform to train and evaluate the performance of Random Forest classifiers on predicting fundraising success. Table~\ref{tab:predictionresults} shows the Random Forest model's accuracy, precision, recall, F-Score, and area under the receiver operating characteristic curve (AUC). On all platforms, the results of the evaluation metrics are strongly correlated. In particular, we achieve accuracy and AUC scores above 0.7. 

\begin{table*}[!ht]
\centering
\caption{Random Forest validation results of predicting fundraising success agree across multiple evaluation metrics. Shown here are the mean 5-fold cross-validation results (all $std\leq0.015$) using 100 estimators over 10,000 iterations and a random under-sampling of the majority class in each iteration.}
\label{tab:predictionresults}
\begin{tabular}{|l|c|c|c|c|c|}
\hline
Category & Accuracy & Precision & Recall & F-Score & AUC \\ \hline
Lending & 0.989 & 0.988 & 0.990 & 0.989 & 0.989 \\
Equity  & 0.882 & 0.886 & 0.876 & 0.881 & 0.882 \\
Charity & 0.691 & 0.720 & 0.626 & 0.670 & 0.691 \\ \hline                  
\end{tabular}
\vspace{-.25cm}
\end{table*}

\begin{figure}[!h]
    \centering
    \includegraphics[scale=.3]{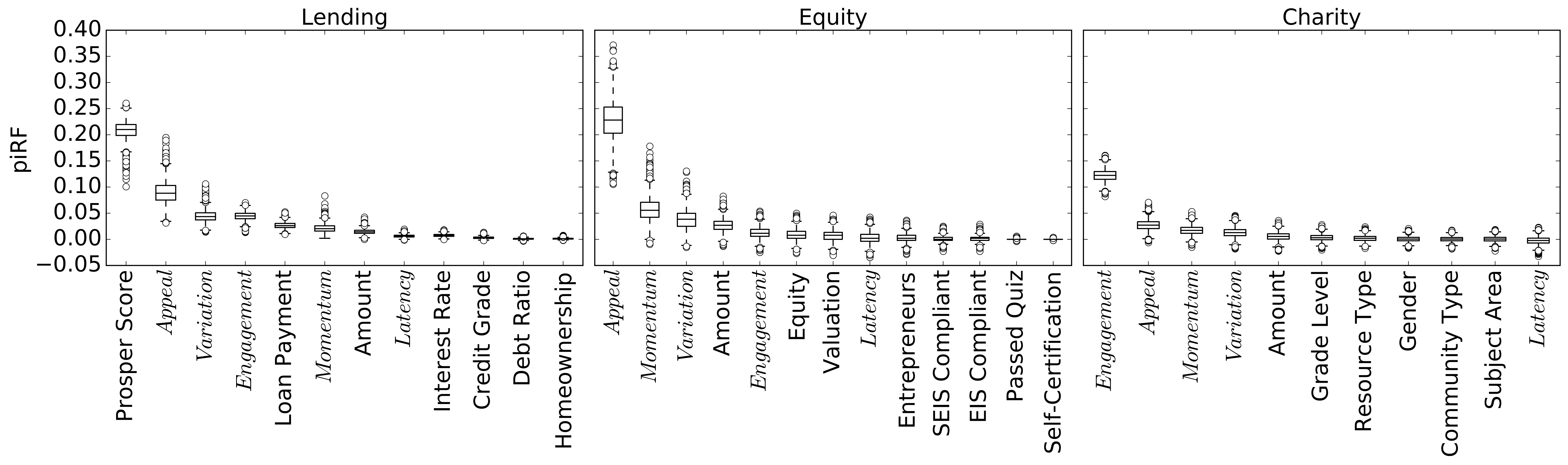}
    \caption{Random Forest permutation importance (piRF) ranking for project and crowd dynamics features. Crowd dynamics features (marked *) account for at least 75\% of the predictive feature importance on all platforms.}
    \label{fig:feature_importance}
    \vspace{-.25cm}
\end{figure}

\begin{figure}[!h]
    \centering
    \includegraphics[scale=.5]{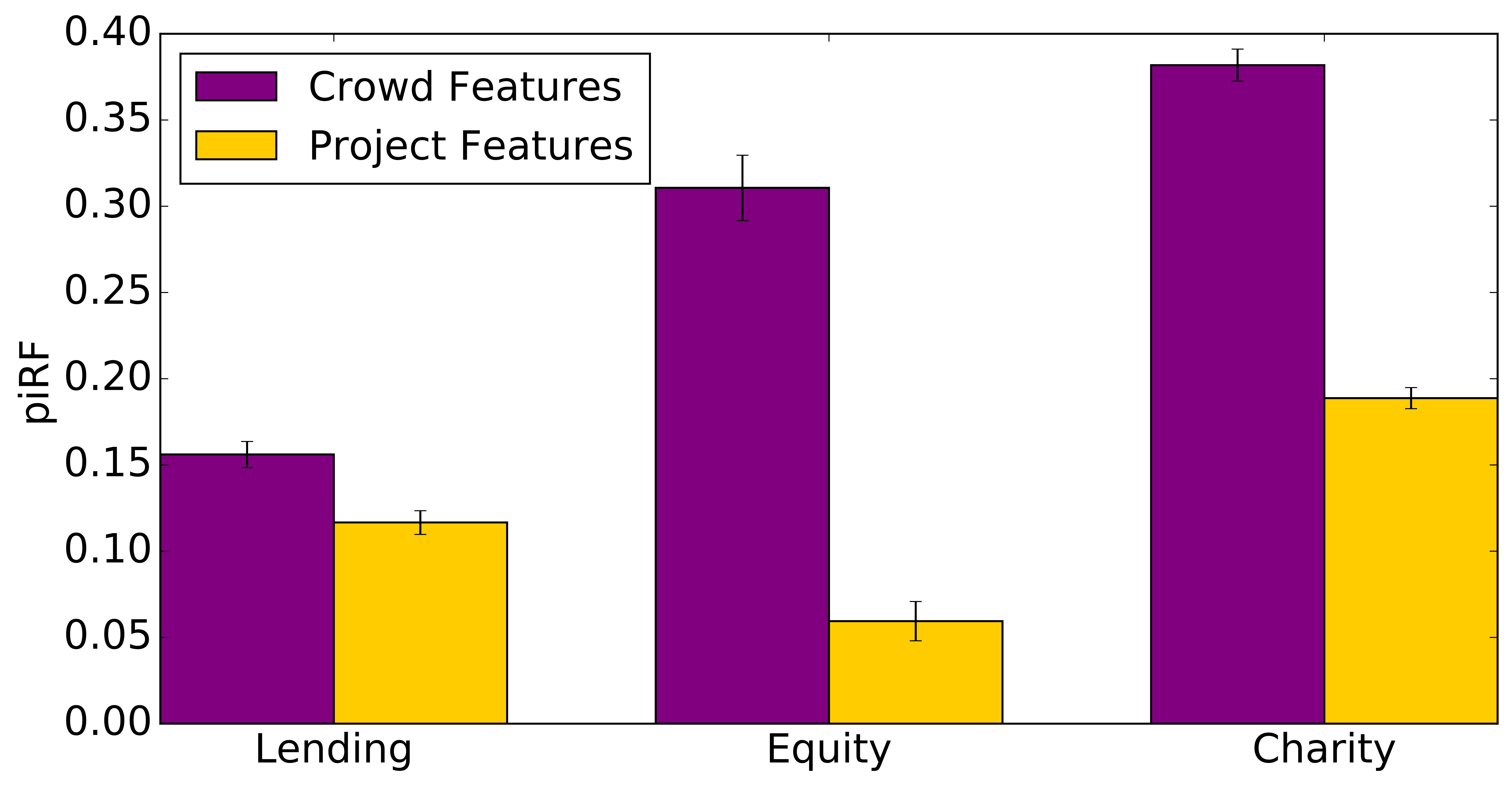}
    \caption{A comparison of the grouped Random Forest permutation importance (piRF) between crowd and project features on all three platforms shows that crowd features are superior to project features in predicting fundraising success.}
    \label{fig:group_piRF}
\end{figure}

Most importantly, we observe across classifiers built for the different platforms that crowd features have relatively higher Random Forest permutation importance (piRF) scores computed on hold-out test sets during cross-validation compared to project features visible to investors, lenders, and donors, respectively. As Figure~\ref{fig:feature_importance} shows, the five crowd features are in the top 7 on the lending platform and top 8 on the equity platform. On the charity platform they occupy the top 4 positions, with latency coming after the project features. Given the simplicity of the latency measure (time difference between first contribution and project posting), unsurprisingly it is the worst-ranked crowd feature across all platforms. Additionally, when grouped together, crowd features account for 57.2\% of the lending, 83.9\% of the equity, and 66.9\% of the charity features' permutation importance (Figure~\ref{fig:group_piRF}). These findings suggest that the dynamics of crowd behaviour add significant value toward predicting fundraising success in crowdfunding, beyond that of traditional project features and further suggest that features deduced from crowd behaviour have huge potential benefits for project creators and crowdfunding platforms (see Section~\ref{sec:discussion}). However, since project features are visible to funders and influence their contribution behaviour, we employed a CEM approach to investigate the causal effects of crowd features irrespective of funders' observations of specific project features.

\subsection{Crowd Features Have Significant Causal Effects towards Fundraising Outcomes}

{To perform CEM, we began by matching funded projects to failed projects with the exact same coarsened project features as explained in Section~\ref{sec:methods}. We matched $7,150$ of $29,013$ funded projects in the lending platform ($L1=0.740$), $198$ of $261$ funded projects in the equity platform ($L1=0.485$), and $1,249$ of $214,531$ funded projects in the charity platform ($L1=0.792$). It is important to highlight that the resulting decrease in the sample sizes of the matched samples is an artefact of matching among only those funded projects for which well matching failed projects exist. From the matched data, we then computed the sample average treatment effect of crowd features on fundraising success. Since the SATT is based on potential outcomes, we interpret the unit-level causal effects in terms of how statistically different they are from zero (no effect) at the 5\% level.}

\begin{figure}[!ht]
    \centering
    \includegraphics[scale=.5]{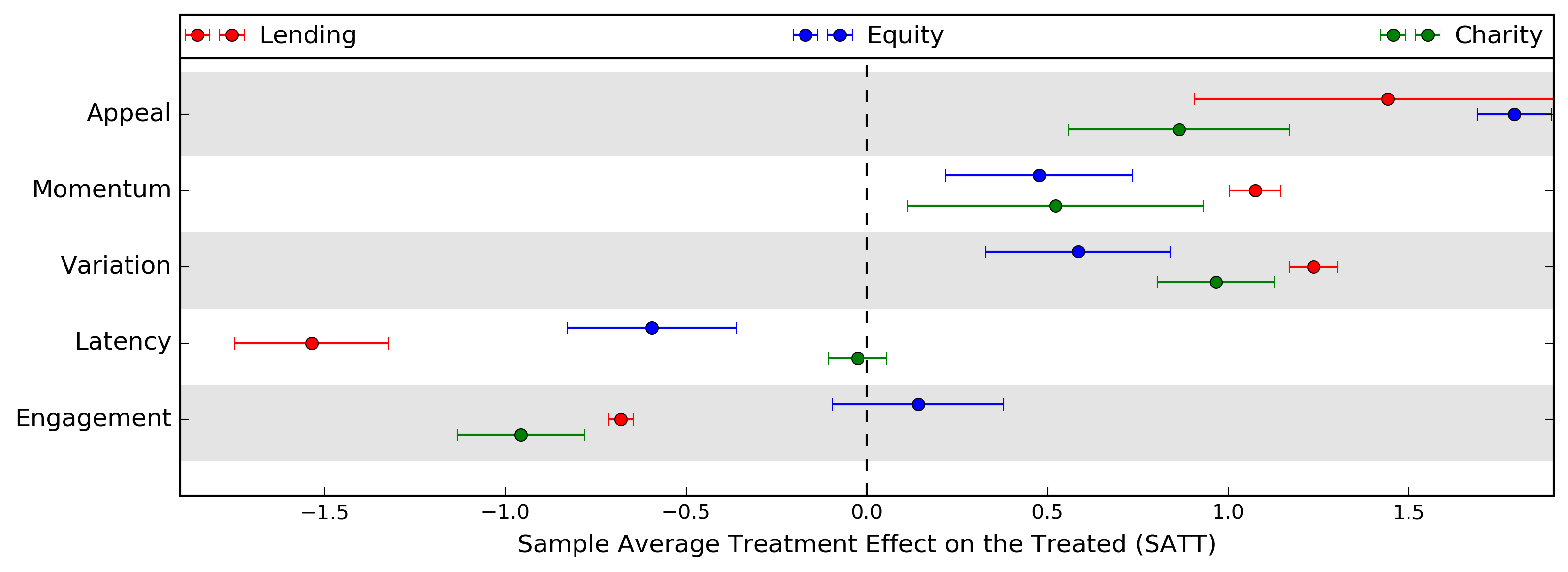}
    \caption{Coarsened Exact Matching (CEM) sample average treatment effect on the treated (SATT) results for the effect of crowd features on fundraising success at 95\% confidence intervals. The SATT estimate is only statistically significant when the 95\% confidence interval (horizontal line) for each crowd feature does not overlap the dotted vertical line at $0$, representing no effect.}
    \label{fig:cem}
\end{figure}

We observed that crowd appeal, momentum, and variation of contributions are significant treatment effects of funding success on all three platforms (Figure~\ref{fig:cem}). Our results show that among projects with similar covariates, some projects may fail to meet their fundraising goal due to low crowd appeal, low momentum, and low variation as well as prolonged latency and engagement. Hence the sooner a project receives funding and the quicker the contributions gain momentum, the more likely the project will be successfully funded independent of its merits. While engagement had a significant effect on fundraising success only on lending and charity platforms, latency had a significant effect on fundraising success only on lending and equity platforms. The treatment effects for both crowd engagement and latency were both negative indicating that the more prolonged the crowd effects, the less chances of project success. These quasi-causal effects further confirm our prior central tendency and correlation results (cf. Table~\ref{tab:crowdstatisticsbyoutcome}) and feature importance results from the Random Forest classifier (cf. Figure~\ref{fig:feature_importance}). Finally, the CEM results further reinforce our Random Forest finding that there are differences in the strength of individual crowd features' association with project outcome. Once again, we find that high appeal, momentum, and variation are robust predictors of fundraising success thereby providing empirical evidence for crowd features that are important indicators of fundraising success across different platforms, while also being robust to the particularities of different online markets.

\section{Discussion}
\label{sec:discussion}
Our work presents a general approach to predicting fundraising success that focuses on the behaviour of the funders rather than the characteristics of project creators or their projects. The presented approach is based on the simple intuition that the timing and amount of funders' contributions have an effect on fundraising outcome. We therefore provide a multi-method analysis for investigating the relationship between the funding crowd's behaviour, as measured using five crowd features, and fundraising success. Through a combination of correlation-based, supervised learning, and quasi-causal inference methods, we demonstrate that our findings regarding the importance of crowd dynamics features in fundraising success are not only stable across different crowdfunding settings, but they are also consistent across three conventional empirical methods. Specifically, we find evidence for the collective nature of success as crowd features are significantly correlated with fundraising success, approximate fundraising success better than the characteristics of projects or their creators, and have significant causal effects towards fundraising outcomes. In the following sections, we elaborate on these findings and their technological implications.

\subsection{The Evolving Nature of Crowdfunding Platforms}
Consistent across three conventional empirical methods, our findings show that the crowd features are robust to the particularities of different crowdfunding platforms and markets, and impartial to platform design and policy changes. This is especially important in studies of crowdfunding due to the evolving nature of both the crowdfunding platforms and markets that make it difficult to consistently investigate the effects of project covariates on fundraising success due to ad-hoc design and policy changes. For example, on the DonorsChoose website, several longitudinal platform changes to location filtering (2004), recommendation (2012), and search (2015) can be expected to influence findings on the effects of both funders' behaviour and project characteristics, such as school location, subject area, and resource type on fundraising success. Specifically, changes in users' ability to filter projects by poverty level (2005), ranking most urgent projects high as the default setting for search (2008), and refining the most urgent criteria to meet both the highest poverty and closest to completion criteria (2012) have been observed in prior literature to increase the effects of project location and community type on fundraising success~\cite{chakraborty2019impact}. In another example, since its SEC registration in 2009, Prosper no longer provides credit grade and other credit information to its prospective lenders. Credit scores, for example, were replaced by the Prosper score which is a custom risk score built using historical in-house data based on Prosper users. Additionally, since 2009, new borrowers to the platform were required to have a FICO score of at least 640, while returning borrowers only needed a score of 600 to request a loan. 

The platform changes identified above affect the type of information presented to funders, the kinds of projects funders are most likely to see, as well as funders' contribution activity. Such platform design and policy changes can be confounding not only when estimating the effect of project features but also when evaluating the impact of crowd behaviour on fundraising success. On the one hand, studies that solely focus on project determinants of fundraising success, i.e. most existing literature on crowdfunding, risk overestimating their findings. For example, platform design features that enable users to filter and search projects by location may increase the importance of projects' location in determining fundraising success compared to platforms that do not afford location search and filtering~\cite{chakraborty2019impact}. On the other hand, studies that focus on crowd-based indicators of fundraising success without controlling for confounding project-level variables risk under-estimating the impact of changes in platform design on crowd behaviour. This is because despite the impact that location search and filtering features, for example, may have on the importance of projects' location in determining fundraising success, these same platform design features may inadvertently impact the crowd appeal of projects of similar quality but different geographic locations. These challenges therefore require controlled approaches to systematically investigate the effects of both project and crowd features on fundraising success on evolving crowdfunding platforms. Our work contributes a framework for studying such scenarios and has implications beyond the study of crowdfunding as well.

\subsection{Main Findings \& Design Contributions}
Through a multi-platform study that aims to improve our understanding of the determinants of fundraising success in different online capital markets, our work engages with ongoing CSCW research on crowdfunding. Specifically, it provides generalisable support for existing empirical and qualitative findings on the role of early contributions~\cite{solomon2015don,colombo2015internal} and presents a suitable approach for controlling for the effects of platform architecture and design changes~\cite{chakraborty2019impact}. Through this approach, we demonstrate the crucial role of three crowd features in determining fundraising success: the crowd appeal, momentum of contributions, and variation in contribution amounts. Prior qualitative work has long emphasised the importance of mobilising a community in crowdfunding, for example by personally reaching out to potential contributors to increase appeal, having an early stage publicity plan to generate fundraising momentum, as well as multiple funding levels (e.g. targeted at big and small funders) to increase the variation in contribution amounts~\cite{hui2014understanding}. Therefore, not only do we lend empirical evidence to the efficacy of mobilising fundraising communities, but we further demonstrate computational approaches for measuring funders' behaviour in terms of the key drivers of fundraising success that characterise different fundraising efforts (i.e. appeal, momentum, and variation). Additionally, these findings support previous qualitative studies that point towards a self–reinforcing pattern whereby early contributions accelerate crowd appeal and momentum through the internal social capital that project creators may develop in the crowdfunding community which in turn provides crucial assistance in igniting a self–reinforcing mechanism that ultimately leads to fundraising success~\cite{colombo2015internal}. Our results further help clarify contradictory findings about the effect of project duration on fundraising success. For instance, our CEM analysis shows that the crowd engagement which corresponds to a project's duration has negative effect on charity and lending platforms. As such, they support the argument that extended activity (i.e. a longer project duration) in crowdfunding settings that rely on small individual contributions may signal the crowd's indecisiveness regarding a project's merits~\cite{mollick2014dynamics,cumming2015crowdfunding,lukkarinen2016success}. At the same time, the positive effect of crowd engagement on fundraising success in the equity platform suggests that when it comes to large capital investments that require significantly more fundraising time and effort (e.g. through due diligence requiring potentially face-to-face interactions in response to higher levels of risk~\cite{agrawal2014some}), longer campaign duration may help to increase the likelihood of project success as the contributions will eventually add up to or even exceed the requested amount~\cite{cordova2015determinants}.

Our findings have important implications for crowdfunding platform design. Having demonstrated that crowd dynamics have significant correlation and causal effects on fundraising outcomes, we believe that the choice architectures of the platforms that mediate crowd behaviour may influence fundraising outcomes. We hope that platform designers can build upon these new and consequential observations to design platforms that harness crowd dynamics in ways that lead to more efficient and successful fundraising. Additionally, our findings are intended to challenge designers to reflect and think more critically about the ways in which their platform design choices enable or inhibit the crowd dynamics that lead to successful fundraising. For instance, how can crowdfunding platforms better signal a project's merit and appeal in such a way that affords funders the ability to quickly and intelligently decide what projects to fund thereby increasing the project's momentum and chances of success? 

We hope that our findings will inspire platform designers to think more broadly about how to create crowdfunding platforms that both promote and support efficient crowd awareness, navigation, and coordination, and are attuned and sensitive to the potential biases and inequalities that may result from inefficient crowd decision-making~\cite{wash2014coordinating}. Our findings also have implications for funders that contribute to these platforms as we show that even for projects of comparable quality, sometimes the difference between funded and not funded is the difference in the funders' behaviour, e.g. whether they find a project appealing, the timing of their contribution, and variation in the amount of their contribution compared to previous contributions. Together, these platform-design and user implications suggest that crowd-aware system design approaches could enhance social navigation and may help to better coordinate crowd behaviour in platform-mediated decision-making environments.

\section{Conclusion}

In this study, we showed that universal features deduced from crowd activity are predictive of fundraising success in different crowdfunding platforms and markets, thereby providing empirical insights on the emergence of collective dynamics that ultimately determine what is worthy of success. Our multi-method analysis has shown that crowd features are correlated with fundraising success, predict fundraising success better than project features, and have a significant effect on fundraising success independent of project features. These results advance a general approach to approximating fundraising success in online capital markets that is robust to platform heterogeneity. Such a general approach is vital considering the evolving nature of crowdfunding platforms both in terms of their user policies and interface design. To better understand how crowdfunding platforms can be designed to promote efficient crowd decision-making, future research should investigate the ways in which the identified crowd features may lead to sub-optimal fundraising outcomes, inefficiencies in capital allocation, or re-enforce existing biases that may exacerbate inequalities. Ultimately, a more nuanced understanding of how crowd behaviour influences fundraising outcomes will inform how crowdfunding and online campaign sites, in general, can be designed to promote the crowd dynamics that lead to successful fundraising to achieve maximal impact.

\section*{Acknowledgments}
This work was supported by the U.S. National Science Foundation under Grant No. IIS-1755873.

\bibliographystyle{unsrt}  
\bibliography{references}  

\end{document}